\documentclass[twocolumn,showpacs,preprintnumbers,amsmath,amssymb]{revtex4}

\usepackage{graphicx}
\usepackage{dcolumn}
\usepackage{bm}

\begin{document}

\title{$\boldsymbol{2n}$-dimensional models with topological mass generation}
%

\author{Shinichi Deguchi} 
\email{deguchi@phys.cst.nihon-u.ac.jp} 
\author{Satoshi Hayakawa}%
\altaffiliation[Present address: ]
{Koito Manufacturing Co., Ltd.}
\affiliation{%
Institute of Quantum Science, College of Science and Technology, 
Nihon University, Chiyoda-ku,  Tokyo 101-8308, Japan
}%


\date{\today}

\begin{abstract}
The 4-dimensional model with topological mass generation 
that has recently been presented 
by Dvali, Jackiw and Pi 
[G. Dvali, R. Jackiw, and S.-Y. Pi, Phys. Rev. Lett. {\bf 96}, 081602 (2006), 
hep-th/0610228] 
is generalized to any even number of dimensions.  
As in the 4-dimensional model, the $2n$-dimensional model describes a mass-generation  
phenomenon due to the presence of the chiral anomaly. 
In addition to this model,  
new $2n$-dimensional models with topological mass generation 
are proposed, in which a St\"{u}ckelberg-type mass term plays a crucial role 
in the mass generation.  
The mass generation of a pseudoscalar field such as the $\eta^{\prime}$ meson is discussed  
within this framework. 
\end{abstract}

\maketitle

\section{\label{sec:level1}Introduction} 
Recently, Dvali, Jackiw and Pi have presented a novel 4-dimensional model 
\cite{DJP} consisting of well-known topological entities: Chern-Pontryagin density  
$\mathcal{P}$ and Chern-Simons current $\mathcal{C}^{\mu}$,  
$\mathcal{P}=\partial_{\mu} \mathcal{C}^{\mu}$.  
This model can describe the mass-generation phenomenon  in a 4-dimensional  
non-Abelian system without treating details of the underlying dynamics.  
Dvali et al. found the model as a partial, 4-dimensional generalization of  
the Schwinger model \cite{Sch} reformulated  
in terms of the topological entities in 2 dimensions. 
The reformulated Schwinger model and the 4-dimensional model 
share the common mass-generation mechanism 
described in topological terms. 
Noting this,   
Dvali et al. stated that the present formulation offers 
a unified topological description of the mass-generation phenomena in seemingly 
unrelated systems.

In this paper, we first consider a straightforward $2n$-dimensional generalization of 
the 4-dimensional model and demonstrate 
that the topological mass generation 
studied by Dvali et al. is present in any even number of dimensions. 
There, as in the 4-dimensional model, it is verified that  
the presence of the chiral anomaly is essential for generating mass.  
Next, we propose a new $2n$-dimensional model with topological mass generation, 
in which a St\"{u}ckelberg-type mass term gives rise to mass generation 
in a gauge invariant manner. 
In addition, we consider a hybrid of the $2n$-dimensional models 
mentioned above, in which a mass is caused by both the St\"{u}ckelberg-type  
mass term and the presence of the chiral anomaly. 
The hybrid model is applied, 
after a few modifications, 
to the mass generation of a pseudoscalar field such as the $\eta^{\prime}$ meson.

In the process of deriving equations of motion in the $2n$-dimensional models, 
it is necessary to know the variation of the Chern-Simons current in $2n$ dimensions.  
To find this, 
we adopt an elegant method developed on $(2n+1)$-dimensional space.

This paper is organized as follows. 
Section 2 introduces the topological entities in $2n$ dimensions.   
Section 3 presents a straightforward $2n$-dimensional generalization of the model found 
by Dvali et al. 
Section 4 proposes new $2n$-dimensional models with a St\"{u}ckelberg-type mass term.  
Section 5 contains a summary and discussion. 
The appendix is devoted to calculating the variation of the Chern-Simons current 
in $2n$ dimensions.

\section{\label{sec:level1}topological entities}

Let $A$ be a (Hermitian) Yang-Mills connection on $2n$-dimensional 
Minkowski space, $M^{2n}$, with local coordinates $(x^{\mu})$.  
The connection $A$ is assumed to take values in a compact  
semisimple Lie algebra $\frak{g}$, and hence $A$ can be expanded as  
$A=gA_{\mu}^{a}T_{a}dx^{\mu}$.  
Here, $g$ is a coupling constant with mass dimension $(2-n)$,  
$\{T_{a} \}$ are Hermitian basis of $\frak{g}$ satisfying 
the commutation relations $[ T_{a}, T_{b} ]=if_{ab}{}^{c} T_{c}$ and the normalization 
conditions ${\rm Tr} (T_{a} T_{b})=\delta_{ab}$. 
The curvature 2-form of $A$ is given by 
\begin{align}
F\equiv dA-i A^2 =\frac{1}{2} g F_{\mu\nu}^{a} T_{a} dx^{\mu} dx^{\nu} , 
\label{1}
\end{align}
with   
$F_{\mu\nu}^{a}=\partial_{\mu} A_{\nu}^{a}-\partial_{\nu} A_{\mu}^{a} 
+g f_{bc}{}^{a} A_{\mu}^{b} A_{\nu}^{c}$. 
(Throughout this paper, the symbol $\wedge$ of the wedge product is omitted.)

Consider the Chern-Pontryagin $2n$-form 
\begin{align}
P_{2n}  & \equiv {\rm Tr} F^{n} 
\nonumber 
\\
&= \frac{1}{2^n} g^{n} h_{a_{1}\cdots a_{n}} 
F^{a_1}_{\mu_{1} \mu_{2}} \cdots F^{a_n}_{\mu_{2n-1} \mu_{2n}} 
\nonumber 
\\ 
& \quad \; \times dx^{\mu_1} dx^{\mu_2} \cdots 
dx^{\mu_{2n-1}} dx^{\mu_{2n}} , 
\label{2}
\end{align}
where $h_{a_{1}\cdots a_{n}}\equiv {\rm Tr}(T_{a_1}\cdots T_{a_{n}})$. 
The Bianchi identity $dF=i(AF-FA)$ guarantees $dP_{2n}=0$. 
Then, in accordance with Poincar\'{e}'s lemma, $P_{2n}$ is expressed  
at least locally as 
\begin{align}
P_{2n}=dC_{2n-1} \,, 
\label{3}
\end{align}
with the Chern-Simons $(2n-1)$-form \cite{CS}
\begin{align}
C_{2n-1}(A, F)\equiv n \int_{0}^{1} dt\, {\rm Tr} (AF_{t}^{n-1}) \,, 
\label{4}
\end{align}
where $F_{t}\equiv tF-i(t^2 -t) A^{2}$.

We now introduce the Hodge $\ast$ operator defined by 
\begin{align}
&\ast( dx^{\mu_{1}} \cdots dx^{\mu_{p}}) 
\nonumber 
\\ 
&=\frac{1}{(2n-p)!} \epsilon^{\mu_{1}\cdots\mu_{p}}{}_{\mu_{p+1}\cdots \mu_{2n}} 
dx^{\mu_{p+1}} \cdots dx^{\mu_{2n}} . 
\label{5}
\end{align}
The $\ast$ operator transforms $p$-forms into their dual $(2n-p)$-forms.  
For a $p$-form 
$\alpha_p=(p!)^{-1}\alpha_{\mu_1 \cdots \mu_p} dx^{\mu_1}\cdots dx^{\mu_p}$ 
on $M^{2n}$, it is verified that  
\begin{align}
\ast \ast \alpha_{p} &=(-1)^{p(2n-p)+1} \alpha_{p} \,, 
\label{6} 
\\ 
\ast \,d \ast \alpha_{p} &=(-1)^{(p-1)(2n-p)+1} 
\partial^{\mu} \alpha_{\mu \mu_{1} \cdots \mu_{p-1}} 
\nonumber 
\\ 
&  \quad \; \times dx^{\mu_1}\cdots dx^{\mu_{p-1}} .
\label{7}
\end{align}
Using (\ref{5}), the Hodge $\ast$ operation of 
$P_{2n}$ is found to be 
\begin{align}
\mathcal{P}_{2n} \equiv \ast P_{2n}
&=\frac{1}{2^{n}} g^n 
h_{a_{1}\cdots a_{n}}  
\epsilon^{\mu_{1}\mu_{2}\cdots\mu_{2n-1}\mu_{2n}} 
\nonumber 
\\ 
& \quad \; \times 
F_{\mu_{1} \mu_{2}}^{a_{1}}  \cdots F_{\mu_{2n-1} \mu_{2n}}^{a_{n}} \,. 
\label{8}
\end{align}
The $0$-form $\mathcal{P}_{2n}$ is referred to as the Chern-Pontryagin density. 
Applying the $\ast$ operator to (\ref{3}) and using the formulas (\ref{6}) and (\ref{7}),  
we have the dual form of (\ref{3}):  
\begin{align}
\mathcal{P}_{2n} =\partial_{\mu} \mathcal{C}_{2n}^{\mu}\, , 
\label{9}
\end{align}
where the $\mathcal{C}_{2n}^{\mu}$ are the components of the 1-form  
$\mathcal{C}_{2n}\equiv -\ast C_{2n-1}$.   
This 1-form, or simply $\mathcal{C}_{2n}^{\mu}$,  
is referred to as the Chern-Simons current. 
The $\mathcal{P}_{2n}$ and $\mathcal{C}_{2n}^{\mu}$  
are topological entities essential for constructing the $2n$-dimensional models  
with topological mass generation.

\section{\label{sec:level1}
Mass Generation Due to Chiral Anomaly}   

Now, we show that the mass-generation mechanism studied in Ref.~\onlinecite{DJP} 
works in any even number of dimensions. 
The Lagrangian that we adopt, $\mathcal{L}_{2n}$, is a $2n$-dimensional analogue of the 
Lagrangian for the 4-dimensional model: 
\begin{align}
\mathcal{L}_{2n}=\frac{1}{2} \mathcal{P}_{2n}^{2} 
+\Lambda^{2} (\mathcal{C}_{2n}^{\nu} -\partial_{\mu} p^{\mu\nu}) 
(\mathcal{J}_{\nu}^{5} -\partial^{\rho} q_{\rho\nu}) \,. 
\label{10} 
\end{align}
Here, $p^{\mu\nu}$ and $q_{\mu\nu}$ are antisymmetric tensor fields, 
$\mathcal{J}_{\nu}^{5}$ is an axial vector current, and $\Lambda$ is a constant 
with mass dimension. (An overall dimensionful constant is omitted.)

Under the (infinitesimal) gauge transformation 
\begin{align}
\delta_{\omega} A_{\mu}^{a} = D_{\mu} \omega^{a} , 
\label{11}
\end{align}
the Chern-Pontryagin density $\mathcal{P}_{2n}$ remains invariant, 
while the Chern-Simons current $\mathcal{C}_{2n}^{\mu}$ transforms as 
\begin{align}
\delta_{\omega} \mathcal{C}_{2n}^{\nu} 
=\partial_{\mu} \mathcal{U}_{2n}^{\mu\nu} \,. 
\label{12}
\end{align}
Here, $\mathcal{U}_{2n}^{\mu\nu}$ is an antisymmetric tensor that is 
a polynomial in $(A_{\mu}^{a}, F_{\mu\nu}^{a}, \omega^{a})$ and linear 
in $\omega^{a}$.  (For further details, see the appendix.)  
We impose the gauge transformation rule 
\begin{align}
\delta_{\omega} p^{\mu\nu} = \mathcal{U}_{2n}^{\mu\nu} 
\label{13} 
\end{align}
on $p^{\mu\nu}$ so that the combination  
$\mathcal{C}_{2n}^{\nu} -\partial_{\mu} p^{\mu\nu}$ 
can be gauge invariant;  
thereby the gauge invariance of $\mathcal{L}_{2n}$ can be secured. 
In this sense, $p^{\mu\nu}$ plays the role of the St\"{u}ckelberg field. 
By contrast, $q_{\mu\nu}$ is assumed to be gauge invariant, 
$\delta_{\omega} q_{\mu\nu}=0$,  by considering  
the gauge invariance of $\mathcal{J}_{\nu}^{5}$. 
As a result, $\mathcal{L}_{2n}$ remains invariant 
under the gauge transformation $\delta_{\omega}$.  
The field $p^{\mu\nu}$ is necessary for the gauge invariance of 
$\mathcal{L}_{2n}$, 
while $q_{\mu\nu}$ is necessary to avoid the integrability condition 
$\partial_{\mu} \mathcal{J}_{\nu}^{5} =\partial_{\nu} \mathcal{J}_{\mu}^{5}$.

As can be seen in the appendix, the variation of the Chern-Simons current 
$\mathcal{C}_{2n}^{\mu}$ is given by (see (\ref{A23})) 
\begin{align}
\delta \mathcal{C}_{2n}^{\nu} 
&=\mathcal{W}_{2n,a}^{\mu\nu} \delta A^{a}_{\mu} 
+\partial_{\mu} \mathcal{V}_{2n}^{\mu\nu} \, , 
\label{14}
\end{align}
where 
\begin{align}
\mathcal{W}_{2n,a}^{\mu\nu} &\equiv 
\frac{n}{2^{n-1}}  g^{n} h
_{a_{1}\cdots a_{n-1} a}  
\epsilon^{\mu_{1}\mu_{2}\cdots \mu_{2n-3} \mu_{2n-2} \mu\nu} 
\nonumber 
\\ 
& \quad \; \times 
F_{\mu_{1} \mu_{2}}^{a_{1}}  \cdots F_{\mu_{2n-3} \mu_{2n-2}}^{a_{n-1}} \,, 
\label{15}
\end{align}
and $\mathcal{V}_{2n}^{\mu\nu}$ is an antisymmetric tensor that is 
a polynomial in $(A_{\mu}^{a}, F_{\mu\nu}^{a}, \delta A^{a}_{\mu})$ and linear 
in $\delta A^{a}_{\mu}$.  Using (\ref{14}), variation of  
the action $S_{2n}=\int \mathcal{L}_{2n} dx$ 
with respect to $A^{a}_{\mu}$ is readily calculated,  
yielding the equation of motion 
\begin{align}
& \{ -\partial_{\mu} \mathcal{P}_{2n} +\Lambda^2 
(\mathcal{J}_{\mu}^5 -\partial^{\rho} q_{\rho\mu} ) \} 
\mathcal{W}_{2n,a}^{\sigma\mu} 
\nonumber 
\\ 
& -\Lambda^2 \partial_{\mu} (\mathcal{J}_{\nu}^5 -\partial^{\rho} q_{\rho\nu} ) 
\frac{\delta \mathcal{V}^{\mu\nu}_{2n}}{\delta A^{a}_{\sigma}} =0 \,. 
\label{16}
\end{align} 
Variation of $S_{2n}$ with respect to $p^{\mu\nu}$ and $q_{\mu\nu}$ yields    
the equations 
\begin{align}
\partial_{\mu} (\mathcal{J}_{\nu}^5 -\partial^{\rho} q_{\rho\nu} ) 
- (\mu \leftrightarrow \nu ) &=0 \,,
\label{17}
\\
\partial^{\mu} (\mathcal{C}^{\nu}_{2n} -\partial_{\rho} p^{\rho\nu} ) 
- (\mu \leftrightarrow \nu ) &=0 \,. 
\label{18}
\end{align}
By virtue of  (\ref{17}), the second line of  (\ref{16}) vanishes.  
Also, we can strip away $\mathcal{W}_{2n,a}^{\sigma\mu}$ in (\ref{16}) 
using the identity  
$\mathcal{W}_{2n,a}^{\sigma\mu} F_{\sigma\nu}^a =2\delta^{\mu}{}_{\nu} \mathcal{P}_{2n}$. 
As a result, provided $\mathcal{P}_{2n} \neq 0$, (\ref{16}) reduces to   
\begin{align}
 -\partial_{\mu} \mathcal{P}_{2n} +\Lambda^2 
(\mathcal{J}_{\mu}^5 -\partial^{\rho} q_{\rho\mu} )=0 \,.
\label{19}
\end{align}
Taking the divergence of  (\ref{19}) and considering antisymmetry of 
$q_{\rho\mu}$, we have  
\begin{align}
\partial^{2} \mathcal{P}_{2n} -\Lambda^2 \partial^{\mu} \mathcal{J}_{\mu}^5 =0 \,. 
\label{20}
\end{align}

Now, we expect that the axial vector current possesses an anomalous divergence: 
\begin{align}
\partial^{\mu} \mathcal{J}_{\mu}^5 =-N \mathcal{P}_{2n} \,, 
\label{21}
\end{align}
where $N$ is a dimensionless positive constant. Then, (\ref{20}) becomes 
\begin{align}
\partial^{2} \mathcal{P}_{2n} +N \Lambda^{2} \mathcal{P}_{2n} =0 \,. 
\label{22}
\end{align}
This shows that the pseudoscalar $\mathcal{P}_{2n}$ has acquired 
the mass $\sqrt{N} \Lambda$. 
It should be stressed that the mass $\sqrt{N} \Lambda$ is generated owing to the presence of 
the chiral anomaly. 
The topological mass generation studied by Dvali et al. \cite{DJP} is thus valid in  
any even number of dimensions.

\section{\label{sec:level1}Other models} 

Until now, we have merely considered a $2n$-dimensional  
generalization of  the 4-dimensional model given in Ref.~\onlinecite{DJP}. 
In this section, we propose new $2n$-dimensional models 
with topological mass generation.

\subsection{\label{sec:level2}A St\"{u}ckelberg-type model} 

With the topological entities $\mathcal{P}_{2n}$ and $\mathcal{C}_{2n}^{\mu}$ 
and the antisymmetric tensor field $p^{\mu\nu}$, 
we first propose a model governed by the Lagrangian 
\begin{align}
\tilde{\mathcal{L}}_{2n}=\frac{1}{2} \mathcal{P}_{2n}^{2} 
-\frac{1}{2} m^{2} (\mathcal{C}_{2n}^{\nu} -\partial_{\mu} p^{\mu\nu}) 
(\mathcal{C}_{2n,\nu} -\partial^{\rho} p_{\rho\nu}) \,, 
\label{23} 
\end{align}
where $m$ is a constant with mass dimension.  
Obviously, $\tilde{\mathcal{L}}_{2n}$ is invariant under the gauge transformation 
$\delta_{\omega}$.

Variation of  the action $\tilde{S}_{2n}=\int \tilde{\mathcal{L}}_{2n} dx$  
with respect to $A_{\mu}^{a}$ gives, with the help of (\ref{14}), 
the equation of motion  
\begin{align}
& \{ -\partial_{\mu} \mathcal{P}_{2n} -m^2  
(\mathcal{C}_{2n,\mu} -\partial^{\rho} p_{\rho\mu} ) \} 
\mathcal{W}_{2n,a}^{\sigma\mu} 
\nonumber 
\\ 
& +m^2 \partial_{\mu} (\mathcal{C}_{2n,\nu} -\partial^{\rho} p_{\rho\nu} ) 
\frac{\delta \mathcal{V}^{\mu\nu}_{2n}}{\delta A^{a}_{\sigma}} =0 \,. 
\label{24}
\end{align} 
Variation of $\tilde{S}_{2n}$ with respect to $p^{\mu\nu}$ yields     
the equation 
\begin{align}
\partial_{\mu} (\mathcal{C}_{2n, \nu} -\partial^{\rho} p_{\rho\nu} ) 
- (\mu \leftrightarrow \nu ) =0 \,. 
\label{25}
\end{align}
By virtue of (\ref{25}), the second line of (\ref{24}) vanishes.  
Also, we can strip away $\mathcal{W}_{2n,a}^{\sigma\mu}$ in (\ref{24}) 
in the same manner as what we used under (\ref{18}). 
Consequently, provided $\mathcal{P}_{2n} \neq 0$,  (\ref{24}) reduces to   
\begin{align}
-\partial_{\mu} \mathcal{P}_{2n} -m^2  
(\mathcal{C}_{2n,\mu} -\partial^{\rho} p_{\rho\mu} )=0 \,.
\label{26}
\end{align}
Taking the divergence of (\ref{26}), and noting (\ref{9}) and 
antisymmetry of $p_{\rho\mu}$, we have  
\begin{align}
\partial^{2} \mathcal{P}_{2n} +m^2 \mathcal{P}_{2n}=0 \,. 
\label{27}
\end{align} 
This shows that the pseudoscalar $\mathcal{P}_{2n}$ has the mass $m$, 
which is immediately caused by the second term on the right-hand side of (\ref{23}). 
Because this term provides a mass in a gauge invariant manner, 
it can be called the St\"{u}ckelberg-type mass term of $\mathcal{P}_{2n}$. 
Accordingly, we refer to the present model as the St\"{u}ckelberg-type model. 
The mass-generation mechanism in this model is obviously different from 
that in the model presented in section 3.

\subsection{\label{sec:level1}A hybrid model} 

Next, we propose a hybrid of the previous two models.  
The Lagrangian that we adopt to define the hybrid is 
\begin{align}
\hat{\mathcal{L}}_{2n}&= \frac{1}{2} \mathcal{P}_{2n}^{2} 
-\frac{1}{2} m^{2} (\mathcal{C}_{2n}^{\nu} -\partial_{\mu} p^{\mu\nu}) 
(\mathcal{C}_{2n,\nu} -\partial^{\rho} p_{\rho\nu}) 
\nonumber 
\\
& \quad \,
+\Lambda^{2} (\mathcal{C}_{2n}^{\nu} -\partial_{\mu} p^{\mu\nu}) 
(\mathcal{J}_{\nu}^{5} -\partial^{\rho} q_{\rho\nu}) \,. 
\label{28} 
\end{align}
This certainly inherits characteristics of the Lagrangians  
(\ref{10}) and (\ref{23}). 
Variation of  the action $\hat{S}_{2n}=\int \hat{\mathcal{L}}_{2n} dx$  
with respect to $A_{\mu}^{a}$ gives the equation of motion  
\begin{align}
& \{ -\partial_{\mu} \mathcal{P}_{2n} -m^2  
(\mathcal{C}_{2n,\mu} -\partial^{\rho} p_{\rho\mu} ) 
\nonumber 
\\
& 
+\Lambda^2 
(\mathcal{J}_{\mu}^5 -\partial^{\rho} q_{\rho\mu} ) \} 
\mathcal{W}_{2n,a}^{\sigma\mu} 
\nonumber 
\\ 
&
 + \{ m^2 \partial_{\mu} (\mathcal{C}_{2n,\nu} -\partial^{\rho} p_{\rho\nu} ) 
\nonumber 
\\
& 
-\Lambda^2 \partial_{\mu} (\mathcal{J}_{\nu}^5 -\partial^{\rho} q_{\rho\nu} ) \} 
\frac{\delta \mathcal{V}^{\mu\nu}_{2n}}{\delta A^{a}_{\sigma}}
=0 \,. 
\label{29}
\end{align} 
Variation of $\hat{S}_{2n}$ with respect to $p^{\mu\nu}$ and $q_{\mu\nu}$ yields    
the equations 
\begin{align}
& m^2 \partial_{\mu} (\mathcal{C}_{2n, \nu} -\partial^{\rho} p_{\rho\nu} ) 
-\Lambda^2 \partial_{\mu} (\mathcal{J}_{\nu}^5 -\partial^{\rho} q_{\rho\nu} ) 
\nonumber  
\\  
& - (\mu \leftrightarrow \nu )=0 \,,
\label{30}
\\
& \partial^{\mu} (\mathcal{C}^{\nu}_{2n} -\partial_{\rho} p^{\rho\nu} )  
- (\mu \leftrightarrow \nu )=0 \,. 
\label{31}
\end{align}
Combining (\ref{30}) and (\ref{31}) leads to (\ref{17}).  
In the same procedure as what was taken 
to derive (\ref{20}) and (\ref{27}) from (\ref{16}) and (\ref{24}), respectively, 
we obtain, from (\ref{29}) and (\ref{30}),  
\begin{align}
\partial^{2} \mathcal{P}_{2n} +m^2 \mathcal{P}_{2n} 
-\Lambda^2 \partial^{\mu} \mathcal{J}_{\mu}^5 =0 \,. 
\label{32}
\end{align} 

When the chiral anomaly is presented, (21) holds and (32) becomes 
\begin{align}
\partial^{2} \mathcal{P}_{2n} 
+(m^2 +N\Lambda^2)  \mathcal{P}_{2n}=0 \,. 
\label{33}
\end{align} 
This demonstrates that the pseudoscalar $\mathcal{P}_{2n}$ has the mass 
$\hat{m}\equiv \sqrt{m^2 +N\Lambda^2}$. 
Obviously, the mass $\hat{m}$ is caused by both the 
St\"{u}ckelberg-type mass term and the presence of the chiral anomaly. 
The hybrid model can be reduced to either of the previous models  
depending on choices of the mass parameters $m$ and $\Lambda$.

\section{\label{sec:level1}Summary and discussion} 

The topological mass generation studied by Dvali et al. is 
valid in any even number of dimensions with no essential changes. 
That is, the $2n$-dimensional 
Chern-Pontryagin density $\mathcal{P}_{2n}$ acquires a mass owing to 
the presence of the chiral anomaly. 
Here, just as in the 4-dimensional model, the presence of the chiral anomaly  
is assumed without specifying its dynamical origin. 
To bring the $2n$-dimensional model close to a complete one,  
it will be necessary to investigate the underlying dynamics that leads to 
the mass generation due to the chiral anomaly.

By incorporating the St\"{u}ckelberg-type mass term into 
the Lagrangian (\ref{10}), 
the $2n$-dimensional model is extended to the hybrid model governed by 
the Lagrangian (\ref{28}).  The hybrid model becomes   
the St\"{u}ckelberg-type model in the absence of the chiral anomaly. 
Now we concentrate our discussion on the hybrid model, because it involves the other 
two models. In the case $n=1$, the hybrid model reduces to   
the 2-dimensional massive Yang-Mills theory with a vector current.

In the case $n\geq2$,  the Lagrangian (\ref{28}) consists of higher dimensional 
terms such as $\mathcal{P}^{2}_{2n}$.  
For this reason, (\ref{28}) cannot be regarded as a fundamental Lagrangian; 
 (\ref{28}) should be viewed as an effective Lagrangian 
that is derived from a fundamental gauge theory. 
The hybrid model in the case $n\geq2$ will be applied to a phenomenological 
description of mass-generation phenomena expected in the fundamental theory. 
In this connection, now we propose an application of the hybrid model to 
the mass generation of a pseudoscalar field. 

As in Ref.~\onlinecite{DJP},  we   
consider the axial vector current of the form 
\begin{align}
\mathcal{J}^{5}_{\mu}=\sqrt{N} \Lambda^{-1} \partial_{\mu} \eta_{0}\,,
\label{34}
\end{align}
where $\eta_{0}$ is a pseudoscalar field. 
Adding an $\eta_{0}$ kinetic term to (\ref{28}), and 
removing $q_{\rho\nu}$ and a total derivative,  
we have the Lagrangian 
\begin{align}
\hat{\mathcal{L}}_{2n}^{\prime}&= \frac{1}{2} \mathcal{P}_{2n}^{2} 
-\frac{1}{2} m^{2} (\mathcal{C}_{2n}^{\nu} -\partial_{\mu} p^{\mu\nu}) 
(\mathcal{C}_{2n,\nu} -\partial^{\rho} p_{\rho\nu}) 
\nonumber 
\\
& \quad \,
-\sqrt{N} \Lambda \mathcal{P}_{2n} \eta_{0} 
+\frac{1}{2} \partial_{\mu} \eta_{0} \partial^{\mu} \eta_{0} \,. 
\label{35} 
\end{align}
This is gauge invariant and leads to the field equations 
\begin{align}
-\partial_{\mu} \mathcal{P}_{2n} -m^2  
(\mathcal{C}_{2n,\mu} -\partial^{\rho} p_{\rho\mu} ) 
+\sqrt{N} \Lambda \partial_{\mu} \eta_{0}=0 \,, 
\label{36}
\\
\partial^{2} \eta_{0} + \sqrt{N} \Lambda \mathcal{P}_{2n}=0 \,, 
\label{37}
\end{align}
and (31). 
Because the divergence of (\ref{34}) reproduces (\ref{21}) with the help of (\ref{37}), 
the chiral anomaly is considered in the Lagrangian (\ref{35}). 
Taking the divergence of (\ref{36}) and using (\ref{9}) and (\ref{37}) yield 
(\ref{33}). Hence, as before, $\mathcal{P}_{2n}$ acquires the mass $\hat{m}$.  
Using (\ref{37}), (\ref{33}) can be written in terms of $\eta_{0}$:  
\begin{align} 
(\partial^{2} +\hat{m}^{2}) \partial^{2} \eta_{0}=0 \,.
\label{38}
\end{align}
This equation implies that $\eta_{0}$ possesses both the massless and massive modes.  
Because the massive mode is recognized to be physical, it follows that 
$\eta_{0}$ can behave as a pseudoscalar field with the mass $\hat{m}$ \cite{Deg}. 
In this way, a mass of the field $\eta_{0}$ is generated.

The Lagrangian (\ref{35}) in 4 dimensions, $\hat{\mathcal{L}}_{4}^{\prime}$, is 
very similar to what Di Vecchia used for solving the U(1) problem in 
a simple model \cite{Vec}. 
The similarity can be seen by identifying $\hat{m}$ and $m$ with 
the masses of the singlet and nonsinglet pseudoscalar-mesons, respectively. 
(The $\eta'$ mass is evaluated by taking into account the mixing between 
the singlet meson $\eta_{0}$ and a nonsinglet meson.) 
A remarkable difference between $\hat{\mathcal{L}}_{4}^{\prime}$ and 
Di Vecchia's Lagrangian, $\mathcal{L}_{\mathrm D}$, 
is that whereas $\mathcal{L}_{\mathrm D}$ contains the mass term  
$\mathcal{M}\equiv -{1\over2}m^{2} \eta_{0}^{2}$,  
$\hat{\mathcal{L}}_{4}^{\prime}$ does not contain it. 
Instead of $\mathcal{M}$, $\hat{\mathcal{L}}_{4}^{\prime}$ contains  
the St\"{u}ckelberg-type mass term to provide the mass $m$. 
Unlike $\mathcal{M}$, the St\"{u}ckelberg-type mass term 
does not break  the symmetry under a constant shift of $\eta_{0}$. 
In spite of such a difference, the hybrid model should have 
a close connection with the effective Lagrangian approach to the U(1) problem 
\cite{Vec,effect}.


\begin{acknowledgments}
We are grateful to Prof. K. Fujikawa for his encouragement   
and useful comments.  
S. D. thanks Prof. R. Banerjee for fruitful comments. 
The work of S. D. is supported in part by  
the Nihon University Research Grant (No. 06-069). 
\end{acknowledgments}


\appendix* 
\section{\label{sec:level1}Variation of the Chern-Simons current} 

In this appendix, we calculate the variation of the Chern-Simons current 
$\mathcal{C}_{2n}^{\mu}$.  For this purpose,  
we adopt a geometric method developed 
on the product space $M^{2n} \times \Bbb{R}$, 
a direct product of $2n$-dimensional Minkowski space $M^{2n}$ 
and 1-dimensional real space $\Bbb{R}$. 
The exterior derivative in $M^{2n} \times \Bbb{R}$ takes the form  
\begin{align}
\boldsymbol{d} =d+\delta_y =\frac{\partial}{\partial x^{\mu}} dx^{\mu}
+\frac{\partial}{\partial y} dy\,, 
\label{A1}
\end{align}
where $y$ denotes the coordinate of $\Bbb{R}$. 
We now consider the following Yang-Mills connection defined on  
$M^{2n} \times \Bbb{R}\,$:  
\begin{align}
\boldsymbol{A}=A+\varOmega = gA_{\mu}^{a}T_{a}dx^{\mu}+g\omega^{a}T_{a}dy \,,  
\label{A2}
\end{align}
where $A$ is a 1-form that, at $y=0$, agrees with the connection $A$ that is   
already present in $M^{2n}$. 
The components $(A_{\mu}^{a},\, \omega^{a})$ of $\boldsymbol{A}$ are 
understood to be functions of $(x^{\mu}, y)$. 
The curvature 2-form of $\boldsymbol{A}$ is defined in the manner same as (\ref{1}): 
\begin{align}
\boldsymbol{F} \equiv \boldsymbol{d} \boldsymbol{A} -i\boldsymbol{A}^{2}.  
\label{A3}
\end{align}
Substituting (\ref{A1}) and (\ref{A2}) into (\ref{A3}) and noting the nilpotency 
$dydy=0$, we have 
\begin{align}
\boldsymbol{F}=F+\varXi \,, 
\label{A4} 
\end{align}
with $\varXi \equiv \delta_y A +D\varOmega$.  
Here, $D\varOmega$ is 
the exterior covariant derivative of $\varOmega$:   
$D\varOmega \equiv d\varOmega-i(A\varOmega+\varOmega A)$. 
Obviously, $\varXi$ can be expressed as 
$\varXi=g \xi^{a}_{\mu} T_{a} dy dx^{\mu}$, with $\xi^{a}_{\mu}$ 
being functions of $(x^{\mu}, y)$. 
Now we write the definition of $\varXi$ as 
\begin{align}
\delta_y A = -D\varOmega +\varXi .
\label{A5}
\end{align}
This expression can be read as a transformation rule of $A$. 
In fact, the right-hand side is understood as 
the sum of the (infinitesimal) gauge 
transformation with a parameter $\varOmega$ and 
the shift transformation with a parameter $\varXi$.  
For the sake of convenience, we decompose (\ref{A5})  
into the sum of the two transformation rules:  
\begin{align}
\delta_{\varOmega} A &= -D\varOmega , 
\label{A6}
\\
\delta_{\varXi} A & = \varXi , 
\label{A7}
\end{align}
in such a way that $\delta_y A=\delta_{\varOmega} A+\delta_{\varXi} A$. 
Accordingly, the exterior derivative $\boldsymbol{d}$ is expressed as 
\begin{align}
\boldsymbol{d}=d+\delta_{\varOmega}+\delta_{\varXi} . 
\label{A8}
\end{align}
The transformation rules (\ref{A6}) and (\ref{A7}) can be written in terms of the component 
fields as 
\begin{align}
\delta_{\omega} A_{\mu}^{a} &= D_{\mu} \omega^{a} , 
\label{A9}
\\
\delta_{\xi} A_{\mu}^{a} & = \xi_{\mu}^{a} \,,  
\label{A10}
\end{align}
with $D_{\mu} \omega^{a} \equiv \partial_{\mu} \omega^{a} 
+g f_{bc}{}^{a} A_{\mu}{}^{b} \omega^{c}$. 
Here, $\delta_{\omega}$ and $\delta_{\xi}$ are defined by 
$\delta_{\varOmega}=\delta_{\omega} dy$ and $\delta_{\varXi}=\delta_{\xi} dy$, 
respectively.

Replacing $(A, F)$ in formula (\ref{3}) by $(\boldsymbol{A}, \boldsymbol{F})$,  
we have an analogue of  (\ref{3}) valid in $M^{2n} \times \Bbb{R}\,$:   
\begin{align}
{\rm Tr} \boldsymbol{F}^{n} =\boldsymbol{d}\boldsymbol{C}_{2n-1} \,, 
\label{A11} 
\end{align}
where $\boldsymbol{C}_{2n-1} \equiv C_{2n-1}(\boldsymbol{A}, \boldsymbol{F})$. 
The $(2n-1)$-form $\boldsymbol{C}_{2n-1}$ can be expanded in 
powers of $dy$; by virtue of the nilpotency $dydy=0$, the expansion    
has only a finite number of expansion terms: 
\begin{align}
\boldsymbol{C}_{2n-1} &= C_{2n-1}(A+\varOmega, F+\varXi)   
\nonumber 
\\ 
&= C_{2n-1}(A, F) +U_{2n-1}(A, F, \varOmega) 
\nonumber 
\\
& \quad\, +V_{2n-1}(A, F, \varXi) \,. 
\label{A12} 
\end{align}
Here, $U_{2n-1}$ is first order in $\varOmega$ and includes no $\varXi$, 
while $V_{2n-1}$ is first order in $\varXi$ and includes no $\varOmega$. 
Concrete forms for $U_{2n-1}$ and $V_{2n-1}$ can be found from   
(\ref{4}) and (\ref{A12}).  
Applying $\boldsymbol{d}$ to (\ref{A12}) gives   
\begin{align}
\boldsymbol{d}\boldsymbol{C}_{2n-1}
&=dC_{2n-1} +dU_{2n-1} +dV_{2n-1}
\nonumber 
\\ 
&\quad\, +\delta_{\varOmega}C_{2n-1} +\delta_{\varXi} C_{2n-1} \,. 
\label{A13} 
\end{align}  
Also, the following expansion is valid with (\ref{A4}): 
\begin{align}
{\rm Tr} \boldsymbol{F}^{n} ={\rm Tr} F^{n} +n {\rm Tr} (F^{n-1} \varXi) \,. 
\label{A14}
\end{align}
Substituting (\ref{A13}) and (\ref{A14}) into (\ref{A11}) 
and decomposing the resultant with respect to $\varOmega$ and $\varXi$, 
we have  
\begin{align}
{\rm Tr} F^{n} &=d C_{2n-1} \,, 
\label{A15} 
\\ 
\delta_{\varOmega} C_{2n-1} &=-d U_{2n-1} \,, 
\label{A16} 
\\
\delta_{\varXi} C_{2n-1} &=n {\rm Tr}(F^{n-1} \varXi) -d V_{2n-1} \,. 
\label{A17} 
\end{align}
Equation (\ref{A15}) is identical to (\ref{3}), 
(\ref{A16}) is the (infinitesimal) gauge transformation of $C_{2n-1}$, and 
(\ref{A17}) is the shift transformation of $C_{2n-1}$. 
In this way, the transformation rules of $C_{2n-1}$ have together been derived.

We can write (\ref{A16}) and (\ref{A17}) as 
\begin{align}
\delta_{\omega} C_{2n-1} &=dU_{2n-2} \,, 
\label{A18} 
\\
\delta_{\xi} C_{2n-1} &=n {\rm Tr}(F^{n-1} \xi) +dV_{2n-2} \,, 
\label{A19} 
\end{align}
with $\xi \equiv g \xi_{\mu}^{a} T_{a} dx^{\mu}$.    
Here, $U_{2n-2}$ and $V_{2n-2}$ are $(2n-2)$-forms defined by  
$U_{2n-1}=U_{2n-2} dy$ and $V_{2n-1}=V_{2n-2} dy$, respectively. 
We hereafter treat (\ref{A18}) and (\ref{A19}) as transformation rules 
in $M^{2n}$ by setting $y=0$. 
Applying the $\ast$ operator to (\ref{A18}) and (\ref{A19}) and using 
the formulas (\ref{6}) and (\ref{7}) lead to the dual forms:   
\begin{align}
\delta_{\omega} \mathcal{C}_{2n}^{\nu} 
&=\partial_{\mu} \mathcal{U}_{2n}^{\mu\nu} \, , 
\label{A20}
\\
\delta_{\xi} \mathcal{C}_{2n}^{\nu} 
&=\mathcal{W}_{2n,a}^{\mu\nu} \xi^{a}_{\mu} 
+\partial_{\mu} \mathcal{V}_{2n}^{\mu\nu} \, ,
\label{A21}
\end{align}
where   
\begin{align}
\mathcal{W}_{2n,a}^{\mu\nu} &\equiv 
\frac{n}{2^{n-1}}  g^{n} h
_{a_{1}\cdots a_{n-1} a}  
\epsilon^{\mu_{1}\mu_{2}\cdots \mu_{2n-3} \mu_{2n-2} \mu\nu} 
\nonumber 
\\ 
& \quad \; \times 
F_{\mu_{1} \mu_{2}}^{a_{1}}  \cdots F_{\mu_{2n-3} \mu_{2n-2}}^{a_{n-1}} \,, 
\label{A22}
\end{align}
the $\mathcal{U}_{2n}^{\mu\nu}$ are the components of the 2-form 
$\mathcal{U}_{2n}\equiv -\ast U_{2n-2}$, 
and the $\mathcal{V}_{2n}^{\mu\nu}$ are the components of the 2-form 
$\mathcal{V}_{2n}\equiv -\ast V_{2n-2}$.  
Obviously, $\mathcal{U}_{2n}^{\mu\nu}$, $\mathcal{V}_{2n}^{\mu\nu}$ 
and $\mathcal{W}_{2n,a}^{\mu\nu}$ are antisymmetric tensors.

Because $\xi_{\mu}^{a}$ are arbitrary functions of $x^{\mu}$, 
the shift transformation (\ref{A10}) can be identified with the variation of $A_{\mu}^{a}$. 
Replacing $\xi_{\mu}^{a}$ by the variation $\delta A_{\mu}^{a}$, 
we express (\ref{A21}) in the form of the variation of $\mathcal{C}_{2n}^{\nu}\,$: 
\begin{align}
\delta \mathcal{C}_{2n}^{\nu} 
&=\mathcal{W}_{2n,a}^{\mu\nu} \delta A^{a}_{\mu} 
+\partial_{\mu} \mathcal{V}_{2n}^{\mu\nu} \, , 
\label{A23}
\end{align}
where $\mathcal{V}_{2n}^{\mu\nu}$ here is linear in $\delta A^{a}_{\mu}$. 
Thus, the variation of the Chern-Simons current has been obtained   
using a geometric method.


\begin{thebibliography}{99}
\bibitem{DJP} G. Dvali, R. Jackiw, and S.-Y. Pi, Phys. Rev. Lett. {\bf 96}, 081602 (2006), 
hep-th/0511175; 
R. Jackiw, hep-th/0610228. 
%
\bibitem{Sch} J. Schwinger, Phys. Rev. {\bf 128}, 2425 (1962).  
%
\bibitem{CS} M. Nakahara, {\it Geometry, Topology and Physics}   
(IOP Publishing Ltd, Bristol, 1990);  
R. A. Bertlmann, {\it Anomalies in Quantum Field Theory}  
(Oxford University Press, Oxford, 1996). 
%
\bibitem{Deg} S. Deguchi, in preparation. 
%
\bibitem{Vec} P. Di Vecchia, Phys. Lett. B {\bf 85}, 357 (1979). 
%
\bibitem{effect}
C. Rosenzweig, J. Schechter and C. G. Trahern, Phys. Rev. {\bf D21}, 3388 (1980); 
P. Di Vecchia and G. Veneziano, Nucl Phys. {\bf B171}, 253 (1980); 
K. Kawarabayashi and N. Ohta, Nucl. Phys.  {\bf B175}, 477 (1980); 
P. Nath and R. Arnowitt, Phys. Rev. {\bf D23}, 473 (1981); 
R. Arnowitt and P. Nath, Phys. Rev. {\bf D25}, 595 (1982). 
\end{thebibliography}
\end{document}